\documentclass[sigconf]{acmart}

\usepackage[ruled]{algorithm2e} 
\usepackage{algorithmic}
\usepackage{makecell}
\usepackage{multirow}
\usepackage{rotating}
\usepackage{amsmath}
\usepackage{color}
\usepackage{flushend}

\SetAlFnt{\small}
\SetAlCapFnt{\small}
\SetAlCapNameFnt{\small}
\SetAlCapHSkip{0pt}
\AtBeginDocument{%
  }

\copyrightyear{2025}
\acmYear{2025}
\setcopyright{acmlicensed}\acmConference[MM '25]{Proceedings of the 33rd
ACM International Conference on Multimedia}{October 27--31, 2025}{Dublin,
Ireland}
\acmBooktitle{Proceedings of the 33rd ACM International Conference on
Multimedia (MM '25), October 27--31, 2025, Dublin, Ireland}
\acmDOI{10.1145/3746027.3755180}
\acmISBN{979-8-4007-2035-2/2025/10}

\begin{document}

\title{Learning Arbitrary-Scale RAW Image Downscaling with Wavelet-based Recurrent Reconstruction}

\author{Yang Ren}
\authornote{The authors contributed equally to this work.}
\affiliation{%
  \institution{School of Aeronautics and Astronautics, Sichuan University}
  \city{Chengdu}
  \country{China}
}
\email{renyang@stu.scu.edu.cn}

\author{Hai Jiang}
\authornotemark[1]
\affiliation{%
  \institution{School of Aeronautics and Astronautics, Sichuan University}
  \city{Chengdu}
  \country{China}
}
\email{jianghai@stu.scu.edu.cn}

\author{Wei Li}
\affiliation{%
  \institution{School of Aeronautics and Astronautics, Sichuan University}
  \city{Chengdu}
  \country{China}
}
\email{li.wei@scu.edu.cn}

\author{Menglong Yang}
\authornote{Corresponding authors.}
\affiliation{%
  \institution{School of Aeronautics and Astronautics, Sichuan University}
  \city{Chengdu}
  \country{China}
}
\email{mlyang@scu.edu.cn}

\author{Heng Zhang}
\affiliation{%
  \institution{Xiaomi Inc.}
  \city{Beijing}
  \country{China}
}
\email{kazenokizi@live.com}

\author{Zehua Sheng}
\affiliation{%
  \institution{Xiaomi Inc.}
  \city{Shanghai}
  \country{China}
}
\email{shengzehua@xiaomi.com}

\author{Qingsheng Ye}
\affiliation{%
  \institution{Xiaomi Inc.}
  \city{Shanghai}
  \country{China}
}
\email{yeqingsheng@xiaomi.com}

\author{Shuaicheng Liu}
\authornotemark[2]
\affiliation{%
  \institution{University of Electronic Science and Technology of China}
  \city{Chengdu}
  \country{China}
}
\email{liushuaicheng@uestc.edu.cn}

\renewcommand{\shortauthors}{Ren et al.}

\begin{abstract}
  Image downscaling is critical for efficient storage and transmission of high-resolution (HR) images. Existing learning-based methods focus on performing downscaling within the sRGB domain, which typically suffers from blurred details and unexpected artifacts. RAW images, with their unprocessed photonic information, offer greater flexibility but lack specialized downscaling frameworks. In this paper, we propose a wavelet-based recurrent reconstruction framework that leverages the information lossless attribute of wavelet transformation to fulfill the arbitrary-scale RAW image downscaling in a coarse-to-fine manner, in which the Low-Frequency Arbitrary-Scale Downscaling Module (LASDM) and the High-Frequency Prediction Module (HFPM) are proposed to preserve structural and textural integrity of the reconstructed low-resolution (LR) RAW images, alongside an energy-maximization loss to align high-frequency energy between HR and LR domain. Furthermore, we introduce the Realistic Non-Integer RAW Downscaling (Real-NIRD) dataset, featuring a non-integer downscaling factor of 1.3$\times$, and incorporate it with publicly available datasets with integer factors (2$\times$, 3$\times$, 4$\times$) for comprehensive benchmarking arbitrary-scale image downscaling purposes. Extensive experiments demonstrate that our method outperforms existing state-of-the-art competitors both quantitatively and visually. The code and dataset will be released at https://github.com/RenYangSCU/ASRD.
\end{abstract}



\begin{teaserfigure}
    \centering
    \includegraphics[width=\linewidth]{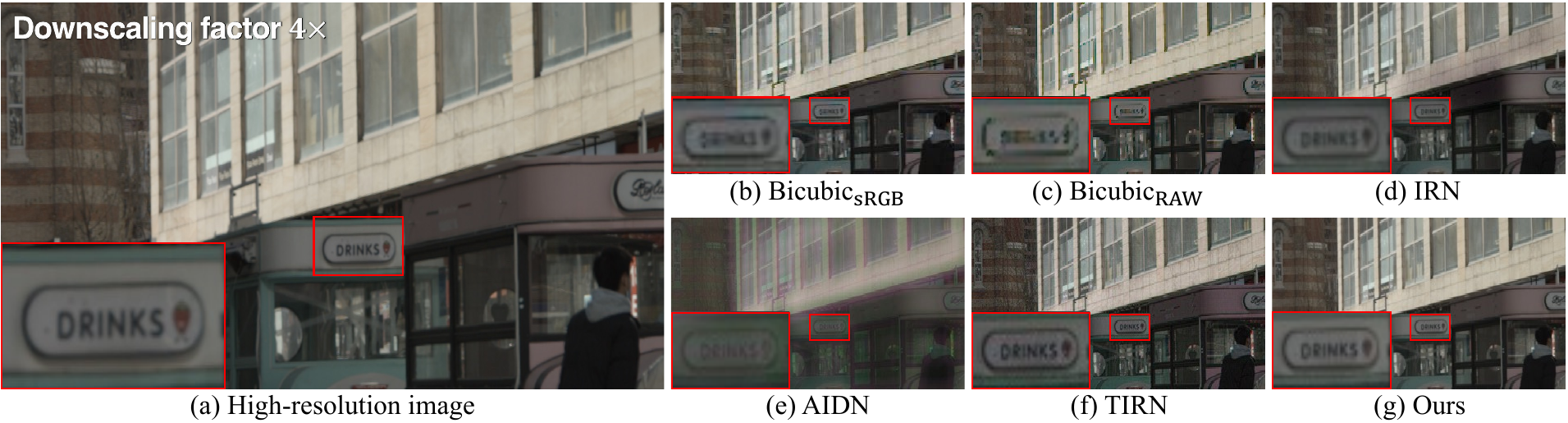}
    \caption{Visual comparisons of the traditional interpolation-based image downscaling method Bicubic that conducted in sRGB and RAW space denoted as Bicubic$_{\text{sRGB}}$ and Bicubic$_{\text{RAW}}$ respectively, learning-based sRGB image downscaling methods IRN~\cite{IRN}, AIDN~\cite{AIDN}, and TIRN~\cite{TSAIN-TIRN}, as well as our proposed method to downscale a high-resolution (HR) image by 4$\times$. For Bicubic$_{\text{RAW}}$ and our method, we employ a fixed ISP pipeline to convert the reconstructed low-resolution RAW images into sRGB format for visualization. Previous methods present blurred details, color distortion, or ghosting artifacts, while our method is capable of generating visually satisfactory results with sharper details.}
    \label{fig: teaser}
\end{teaserfigure}
\maketitle

\section{Introduction}\label{sec:intro}
Modern photographic devices, such as DSLR cameras and smartphones, are capable of capturing images at ultra-definition resolutions to preserve sufficient visual details, while these images inevitably undergo resolution reduction to enable efficient storage, facilitate bandwidth-constrained transmission, and ensure interoperability with diverse mobile displays and web application interfaces~\cite{CAR, TAU}. Image downscaling, therefore, has emerged as a fundamental operation in computational imaging to transform high-resolution (HR) images into low-resolution (LR) images while preserving perceptual fidelity and structural consistency~\cite{L0-regularized, Perceptually}. 

Existing methods predominantly perform downscaling in the sRGB domain, leveraging either interpolation-based techniques (e.g., Bilinear or Bicubic) or learning-based frameworks~\cite{CAR, IARN, IRN, TSAIN-TIRN, IRN-Extension}. Interpolation-based approaches are widely adopted due to their computational efficiency but typically result in blurred details and fine-grained texture loss to degrade the perceptual quality and adversely affect downstream tasks. As shown in Fig.~\ref{fig: teaser}(b), the interpolation-based method Bicubic under a downscaling factor of $4\times$ generates a downscaled image with diminished details. To address these limitations, recent research has increasingly focused on learning-based paradigms to generate visually appealing sRGB images. However, existing learning-based methods typically perform image rescaling pipeline, where downscaling is followed by upscaling, which prioritizes improving the quality of reconstructed high-resolution images from low-resolution counterparts, rather than optimizing the downscaling process itself, thus leading to suboptimal preservation of fine-grained details and textures in the low-resolution representations. As shown in Fig.~\ref{fig: teaser}(d)-(f), previous learning-based methods IRN~\cite{IRN}, AIDN~\cite{AIDN}, and TIRN~\cite{TSAIN-TIRN} present compromised downscaled images with blurred details, color distortion, or ghosting artifacts to degrade the visual quality. 

RAW images, with their unprocessed photonic information captured by camera sensors, offer greater flexibility compared to sRGB images. This capability makes RAW images particularly well-suited for advanced image processing tasks, motivating us to explore image downscaling in the RAW domain. However, to date, there are no advanced learning-based methods specifically tailored for RAW image downscaling. Although traditional interpolation-based techniques can be applied to RAW data, their direct employment often results in jump-sampling artifacts due to the inherent characteristics of RAW images. As shown in Fig.~\ref{fig: teaser}(c), the interpolation-based method Bicubic performed in the RAW domain, denoted as Bicubic$_{\operatorname{RAW}}$, generates a low-resolution image with severe aliasing effects that reduce edge integrity, thereby highlighting its ineffectiveness for RAW image downscaling tasks.

In this paper, we propose a wavelet-based recurrent reconstruction framework for the arbitrary-scale RAW image downscaling task to fully leverage the flexibility of RAW images to generate low-resolution images with satisfactory visual fidelity and perceptibility. Specifically, to address the challenge of directly mapping high-resolution RAW images to low-resolution ones, we leverage the information lossless decomposition property of wavelet transformation to reconstruct the low-resolution RAW image in a coarse-to-fine manner. During the progressive reconstruction process, we introduce a Low-Frequency Arbitrary-Scale Downscaling Module (LASDM) and a High-Frequency Prediction Module (HFPM) to sequentially refine the downscaled low-frequency and high-frequency components. To further preserve high-frequency details in the reconstructed low-resolution images, we propose an energy-maximization loss function $\mathcal{L}_{em}$, which restricts the energy of the low-resolution high-frequency maps to be close to that of the corresponding high-resolution maps. As shown in Fig.~\ref{fig: teaser}(g), our method is capable of generating visually satisfactory low-resolution results with sharper details. Furthermore, we present a \textbf{Real}istic \textbf{N}on-integer \textbf{R}AW \textbf{D}ownscaling (Real-NIRD) dataset that contains realistic paired RAW images with a non-integer downscaling factor of 1.3$\times$, and incorporate it with the Real-RAWVSR~\cite{Real-rawvsr} dataset, which includes integer downscaling factors (2$\times$, 3$\times$, 4$\times$), to enable comprehensive benchmarking arbitrary-scale RAW image downscaling. Overall, our contributions are summarized as follows:
\begin{itemize}
    \item  We propose a wavelet-based recurrent reconstruction downscaling framework that leverages the information lossless property of wavelet transformation for arbitrary-scale RAW image downscaling in a coarse-to-fine manner.
    \item We propose a low-frequency arbitrary-scale downscaling module and a high-frequency prediction module, complemented by an energy-maximization loss, to preserve the structural and textural integrity, facilitating the generation of low-resolution images with satisfactory visual quality.
    \item We introduce the Real-NIRD dataset that contains realistic paired RAW images with a non-integer downscaling factor of 1.3$\times$. Experiments on our proposed Real-NIRD dataset and the publicly available dataset with integer downscaling factors demonstrate the superiority of the proposed method.
\end{itemize}

\section{Related Work}
\label{sec:related_work}
\begin{figure*}
    \centering
    \includegraphics[width=\linewidth]{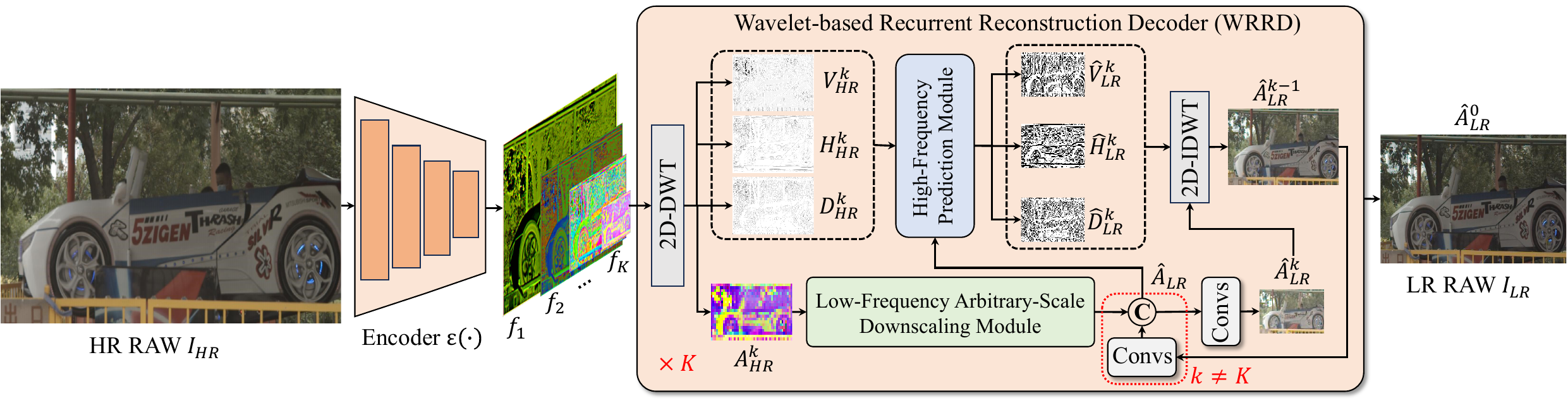}
    \caption{The overall pipeline of our proposed arbitrary-scale RAW image downscaling framework. Our approach first extracts pyramid features $\{f_k\}_{k=1}^{K}$ from the high-resolution (HR) RAW image $I_{HR}$ via an encoder $\mathcal{E(\cdot)}$, and then hierarchically reconstructs the low-resolution (LR) RAW image $I_{LR}$ through the proposed Wavelet-based Recurrent Reconstruction Decoder (WRRD). The WRRD operates across $K$ levels, where the feature $f_{k}$ at each level is transformed into the wavelet domain via 2D discrete wavelet transform (2D-DWT) and separately processes the low-frequency coefficient $A_{HR}^{k}$ and the high-frequency components $\{V_{HR}^{k}, H_{HR}^{k}, D_{HR}^{k}\}$ through the newly proposed Low-Frequency Arbitrary-Scale Downscaling Module (LASDM) and High-frequency Prediction Module (HFPM). Finally, the reconstructed wavelet components recursively undergo 2D inverse discrete wavelet transform (2D-IDWT), generating the final downscaled LR RAW image $\hat{I}_{LR}$.}
    \label{fig: pipeline}
\end{figure*}
\subsection{Image Downscaling and Rescaling} 
Image downscaling, traditionally based on interpolation techniques like bilinear, bicubic~\cite{bicubic}, area, and nearest, forms the backbone of modern image processing. Given the maturity of these traditional approaches, we primarily review recent advances in algorithmic improvements for image downscaling. For instance, spectral remapping~\cite{gastal2017spectral} utilizes discrete Gabor frequency decomposition to transfer high-frequency details into the downsampled image spectrum. $\mathcal{L}_{0}$-regularized~\cite{L0-regularized} integrates a dual L0-regularized optimization scheme to retain essential structures during downscaling.  With the rise of learning-based approaches, CNN-based downscaling~\cite{CNN-CR} has reframed the task as a compact resolution learning task. However, subsequent research has focused more on bi-directional mapping between LR and HR domains rather than solely on downscaling.

Image rescaling is essential for various computer vision applications, especially on edge devices. Several learning-based methods~\cite{TAU, CAR, IRN, IRN-Extension, IARN, TSAIN-TIRN} have been explored to enhance high-resolution (HR) reconstruction. TAU~\cite{TAU} establishes an auto-encoding framework for the joint optimizing downscaling and upscaling networks to improve reconstruction fidelity. CAR~\cite{CAR} further incorporates content-adaptive resamplers co-trained with super-resolution models. IRN~\cite{IRN, IRN-Extension} leveraged invertible neural networks (INNs) to model HR-to-LR transformations as bijective mappings while encoding residual information into latent variables. Despite their effectiveness, these techniques remain constrained to predefined integer scaling ratios (e.g., ×2, ×4). AIDN~\cite{AIDN} overcomes this limitation by enabling arbitrary-scale downscaling with faithful HR reconstruction. Parallel developments include IARN~\cite{IARN}, which achieves similar flexibility through subpixel value functions based on local implicit image representations. TIRN~\cite{TSAIN-TIRN} proposed a plug-and-play tri-branch invertible block that decomposes the low-frequency branch into luminance (Y) and chrominance (CbCr) components. Additionally, IDA-RD~\cite{IDA-RD} has been introduced to assess the quantitative quality of downscaled images as non-reference metrics, framing downscaling and super-solution as the encoding and decoding processes in a rata-distortion model. Overall, image rescaling algorithms typically regard downscaling as an intermediate step and often overlook the quality of the resulting low-resolution images.

\subsection{Arbitrary-Scale Image Super-Resolution} 
Arbitrary-scale image super-resolution algorithms have gained significant attention for their efficiency in real-world applications. Research efforts~\cite{EDSR, meta-sr, ArbSR, SRWarp, LIIF, LTE} aim to overcome the limitations of conventional, scale-specific models. EDSR~\cite{EDSR} merges multiple scale models into a single framework (MDSR). MetaSR~\cite{meta-sr} introduced a unified arbitrary-scale model by incorporating the scale factor as input to predict the upscaling filter. Recently, ArbSR~\cite{ArbSR} introduced a modular plug-in approach with conditional convolutions to generate dynamic, scale-aware filters. SRWarp~\cite{SRWarp} used a differentiable warping layer to handle geometric deformations. While improving flexibility, challenges remain in generalizing to out-of-distribution cases. To address this, LIIF~\cite{LIIF} proposed an encoder that learns continuous image representations with implicit neural representations for handling arbitrary resolutions, and LTE~\cite{LTE} enhanced this with a dominant-frequency estimator. LINF~\cite{LINF} employed normalizing flow to model texture details at various scales, addressing the ill-posed problem for high-quality results. AnySR~\cite{AnySR} transformed arbitrary-scale SR into a more efficient, scalable method with improved performance through feature-interweaving.


\section{Method}\label{sec:method}
\subsection{Overview}\label{subsec: overview}
The overall pipeline of our proposed arbitrary-scale RAW image downscaling framework is illustrated in Fig.~\ref{fig: pipeline}. Given a packed high-resolution (HR) RAW image $I_{HR} \in \mathbb{R}^{H \times W \times 4}$, we first employ a CNN encoder $\mathcal{E}(\cdot)$ which consists of $K$ cascade residual blocks where each block downsamples the input by a scale of $2$ to transform the input image into pyramid features denoted as $f_{k} \in \mathbb{R}^{\frac{H}{2^{k-1}}\times\frac{W}{2^{k-1}}\times c_{k}}, k \in [1, K]$. Then, we introduce a Wavelet-based Recurrent Reconstruction Decoder (WRRD) that leverages the information lossless characteristic of wavelet transformation to effectively preserve both the structural integrity and fine details of the low-resolution (LR) RAW image. The WRRD operates across $K$ levels, where the feature at each level is transformed into the wavelet domain via 2D discrete wavelet transform (2D-DWT) and separately processes the low-frequency coefficient $A_{HR}^{k}$ and the high-frequency components $\{V_{HR}^{k}, H_{HR}^{k}, D_{HR}^{k}\}$ through the newly proposed Low-Frequency Arbitrary-Scale Downscaling Module (LASDM) and High-frequency Prediction Module (HFPM). Finally, the reconstructed wavelet sub-bands recursively undergo 2D inverse discrete wavelet transform (2D-IDWT), ultimately yielding the final downscaled LR RAW image $\hat{I}_{LR} \in \mathbb{R}^{h \times w \times 4}$.

\subsection{Wavelet-based Recurrent Reconstruction}
For RAW image downscaling, it is essential to effectively merge redundant information from HR images while preserving fine detail and maintaining accurate structural representation. However, directly mapping HR images to LR counterparts is non-trivial due to the inherent characteristics of RAW images and the scale-independent nature of the downscaling task. To this end, we propose to utilize the merits of lossless image decomposition with wavelet transformation to predict the multilevel decomposition components of the LR RAW image and then progressively reconstruct the final result in a coarse-to-fine manner.

\textbf{2D Discrete Wavelet Transformation.} Given a packed RAW image $I \in\mathbb{R}^{H \times W \times 4}$, we use 2D-DWT with Haar wavelets~\cite{2D-DWT} to transform it into four sub-bands, i.e.,
\begin{equation}
    \{A^{1}, V^{1}, H^{1}, D^{1}\} = \text{2D-DWT}(I),
\end{equation}
where $A^{1}, V^{1}, H^{1}, D^{1} \in \mathbb{R}^{\frac{H}{2}\times\frac{W}{2}\times 4} $ represent the average of the input image and high-frequency information in the vertical, horizontal, and diagonal directions, respectively. Specifically, the average coefficient represents the global information of the original image, which can be treated as a downsampled version of the image, and the other three coefficients contain sparse local details. Therefore, the target LR RAW image can be regarded as the average coefficient $A_{LR}^{0}$, which can be progressively reconstructed from $A_{LR}^{K}$ with a set of accurate high-frequency information through 2D-IDWT, where $A_{LR}^{K}$ denotes the average coefficient after $K$ times wavelet transformations on $A_{LR}^{0}$, i.e.,
\begin{equation}
    A_{LR}^{k-1} = \text{2D-IDWT}\{A_{LR}^{k},V_{LR}^{k}, H_{LR}^{k}, D_{LR}^{k}\},
\end{equation}
where $\{A_{LR}^{k},V_{LR}^{k}, H_{LR}^{k}, D_{LR}^{k}\}\in\mathbb{R}^{\frac{H}{2^k}\times\frac{W}{2^k}\times c}, k\in[1, K]$. 

Based on this theory, we design a Wavelet-based Recurrent Reconstruction Decoder (WRRD) with $K$ levels, where each level is designed to generate the HR low-frequency and high-frequency maps with target spatial dimensions. The major objective is to preserve the structure of the low-frequency components while maintaining the sharp details of the high-frequency components, a goal that can be more effectively achieved through recursive reconstruction. Specifically, the HR feature map $f_{k}$ is transformed into the wavelet domain denoted as $\{A_{HR}^{k}, V_{HR}^{k}, H_{HR}^{k}, D_{HR}^{k}\}$. Since the redundant information during the downscaling task should primarily be merged in the low-frequency component while high-frequency information should be largely preserved. We, therefore, propose a Low-Frequency Arbitrary-Scale Downscaling Module (LASDM) to abstract the main structure of $A_{HR}^{k}$ to reconstruct the corresponding $\hat{A}_{LR}^{k}$ at the arbitrary scale, and propose a High-Frequency Prediction Module (HFPM) to predict the low-resolution high-frequency components $\{\hat{V}_{LR}^{k}, \hat{H}_{LR}^{k}, \hat{D}_{LR}^{k}\}$ from the predicted $\hat{A}_{LR}^{k}$ and original high-frequency counterparts to serve as supplementary material for assisting in the next low-frequency map generation. Finally, the predicted four wavelet sub-bands are utilized to generate the low-frequency coefficient $\hat{A}_{LR}^{k-1}$ through 2D-IDWT, which is then concatenated with the output of LASDM in the ($k-1$)-th level to predict the low-frequency coefficient in the next iteration. 

\subsection{Proposed Modules}
\begin{figure}[!t]
    \centering
    \includegraphics[width=\linewidth]{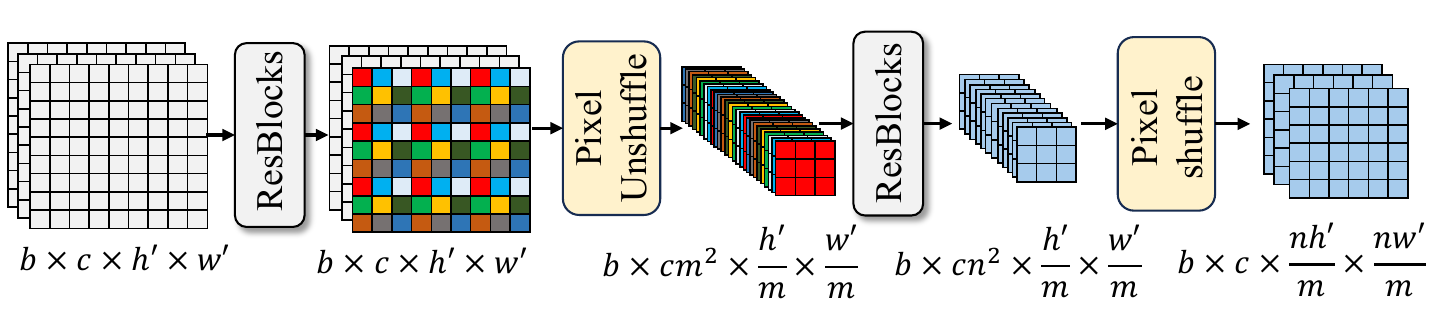}
    \caption{The detailed architecture of our proposed low-frequency arbitrary-scale downscaling module.}
    \label{fig: LASDM}
\end{figure}

\textbf{Low-Frequency Arbitrary-Scale Downscaling Module.} Due to the inherent characteristics of RAW images, the HR-to-LR RAW image mapping is inherently a cross-channel operation, presenting challenges even with low-frequency components. While integer downsampling is relatively straightforward as pixel mappings and receptive fields are clearly defined and spatially aligned, non-integer downscaling (i.e., scaling by arbitrary fractional factors) requires interpolation at arbitrary locations, often causing aliasing and loss of structure. Moreover, non-integer downscaling lacks a consistent and invertible input-output mapping, complicating efficient and consistent sampling strategies without introducing distortion or spatial inconsistency. To address these issues, we propose a Low-frequency Arbitrary-Scale Downscaling Module (LASDM) that leverages the benefits of pixel shuffle and pixel unshuffle operations~\cite{pixel-shuffle} to merge information in the channel dimension for both integer and non-integer downscaling, as shown in Fig.~\ref{fig: LASDM}.

Specifically, whatever integer or non-integer downscaling factors can be represented as a rational fraction $\frac{n}{m}$, enabling its decomposition into a combination of integer downscaling and upscaling operations. Therefore, our LASDM first employs several Res Blocks followed by pixel unshuffle operation with a scale factor of $m$ to transform the input $k$-th level high-resolution low-frequency feature $A_{HR}^{k} \in \mathbb{R}^{h^{\prime} \times w^{\prime} \times c}$, where $h^{\prime} =\frac{H}{2^k}$ and $w^{\prime}=\frac{W}{2^k}$, into low-resolution one with spatial dimension of $\frac{h^{\prime}}{m} \times \frac{w^{\prime}}{m} \times cm^2$. Subsequently, since $m>n$, we leverage several Res Blocks to merge the low-frequency information in channel-wise into the feature with the dimension of $\frac{h^{\prime}}{m} \times \frac{w^{\prime}}{m} \times cn^2$, ensuring the preservation of the core structure of low-frequency information. Finally, the pixel shuffle operation with a scale factor of $n$ is employed to obtain $\hat{A}_{LR} \in \mathbb{R}^{\frac{nh^{\prime}}{m} \times \frac{nw^{\prime}}{m} \times c}$ with target resolution. Overall, our LASDM is capable of handling low-frequency information while preserving the image structure and details for arbitrary-scale downsampling.
\begin{figure}[!t]
    \centering
    \includegraphics[width=0.8\linewidth]{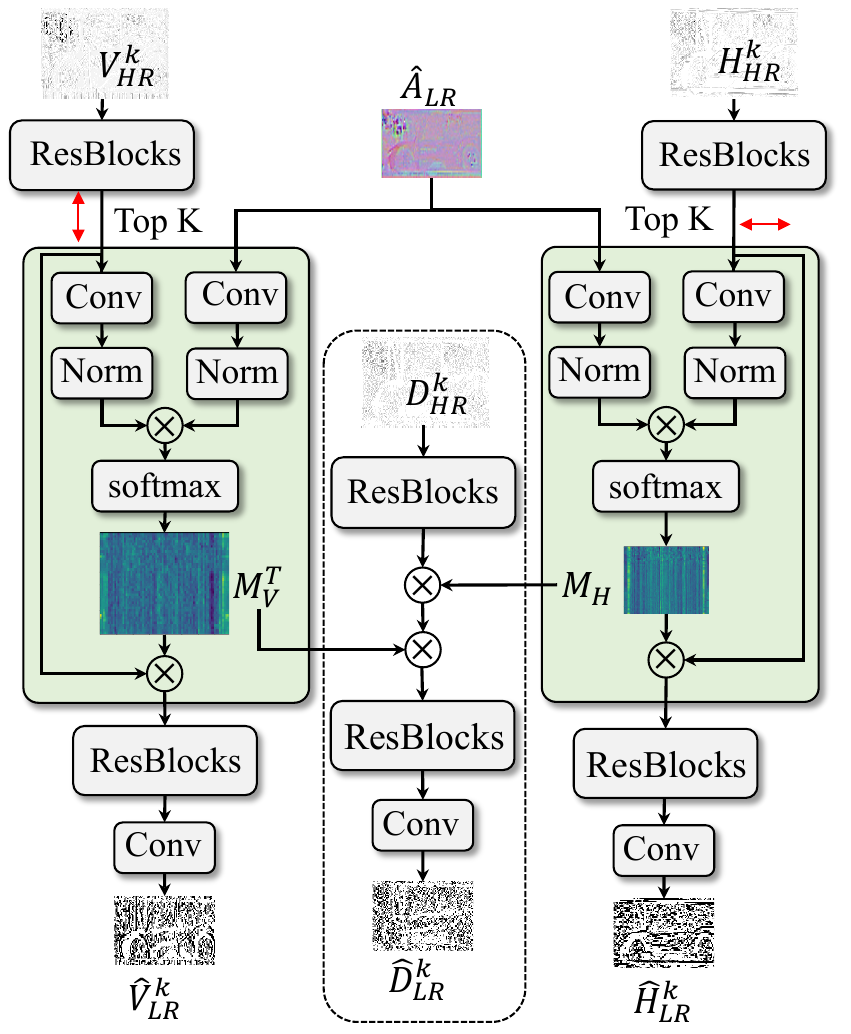}
    \caption{The detailed architecture of our proposed high-frequency prediction module.}
    \label{fig: HFPM}
\end{figure}

\textbf{High-Frequency Prediction Module.} In the $k$-th level, the high-frequency maps in the vertical, horizontal, and diagonal directions, i.e., $\{V_{HR}^{k}, H_{HR}^{k}, D_{HR}^{k}\}$, can serve as complement to the corresponding low-frequency map to preserve the high-frequency details, enabling the lossless reconstruction of $A_{LR}^{k-1}$. To this end, we introduce the High-frequency Prediction Module (HFPM) to predict the corresponding low-resolution high-frequency maps from original high-resolution high-frequency features and the predicted low-resolution low-frequency component. 

To be specific, we first extract the top $\frac{nh^{\prime}}{m}$ values along the vertical direction of $V_{HR}^{k}$ and the top $\frac{nw^{\prime}}{m}$ values along the horizontal direction of $H_{HR}^{k}$ after residual blocks, retaining the most significant sparse high-frequency information, which can be denoted as $V^{\prime} \in \mathbb{R}^{\frac{nh^{\prime}}{m} \times w^{\prime} \times c}$ and $H^{\prime} \in \mathbb{R}^{h^{\prime} \times \frac{nw^{\prime}}{m} \times c}$. Since $H^{\prime}$ lacks precise localization of sparse high-frequency details at low resolution, we leverage a cross-attention module to remap the high-frequency information, yielding the low-resolution horizontal feature $\hat{H}_{LR}^{k} \in \mathbb{R}^{\frac{nh^{\prime}}{m} \times \frac{nw^{\prime}}{m} \times 4}$ along with the location map ${M}_{H} \in \mathbb{R}^{\frac{nh^{\prime}}{m} \times h^{\prime} \times c}$, which is formulated as:
\begin{equation}
    {M}_{H}=\operatorname{softmax} ( \frac{\operatorname{Conv}(\hat{A}_{LR}) \times \operatorname{Conv}(H^{\prime})^T}{||\operatorname{Conv}(\hat{A}_{LR})||_2 \cdot ||\operatorname{Conv}(H^{\prime})||_2}) ,
\end{equation}
\begin{equation}
     \hat{H}_{LR}^{k} = \text{Conv}(\text{ResBlocks}({M}_{H} \times H^{\prime})),
\end{equation}
where $\times$ represent the matrix multiply and $||\cdot||_2$ represents L2 normalization. In parallel, we obtain ${M}_{V}\in \mathbb{R}^{\frac{nw^{\prime}}{m} \times w^{\prime} \times c}$ and $\hat{V}_{LR}^{k} \in \mathbb{R}^{\frac{nh^{\prime}}{m} \times \frac{nw^{\prime}}{m} \times 4} $ in the similar way. Since diagonal information should remain consistent with the other two directions, we employ both vertical and horizontal location maps to achieve the diagonal direction mapping, which can be formulated as:
\begin{equation}
    \hat{D}_{LR}^{k} = \text{Conv}(\text{ResBlocks}({M}_{H} \times D_{HR}^{k} \times {M}_{V}^T )),
\end{equation}
where $\hat{D}_{LR}^{k} \in \mathbb{R}^{\frac{nh^{\prime}}{m} \times \frac{nw^{\prime}}{m} \times 4}$. Moreover, the input features are incorporated as a residual in the independent branch by matching the target size. With the integration of HFPM, the high-frequency details at target scales can be accurately predicted.

\begin{table*}[!t]
  \centering
    \resizebox{\linewidth}{!}{
    \Large
    \begin{tabular}{c|l|ccc|ccc|ccc|ccc}
    \toprule
    \multirow{2}[4]{*}{Type} & \multirow{2}[4]{*}{Method} & \multicolumn{3}{c|}{1.3$\times$} & \multicolumn{3}{c|}{2$\times$} & \multicolumn{3}{c|}{3$\times$} & \multicolumn{3}{c}{4$\times$} \\
    \cmidrule{3-14}          &       & PSNR $\uparrow$ & SSIM $\uparrow$ & LPIPS $\downarrow$ & PSNR $\uparrow$ & SSIM $\uparrow$ & LPIPS $\downarrow$ & PSNR $\uparrow$ & SSIM $\uparrow$ & LPIPS $\downarrow$ & PSNR $\uparrow$ & SSIM $\uparrow$ & LPIPS $\downarrow$ \\
    \midrule
    \multirow{8}[2]{*}{\rotatebox{90}{sRGB-based}} 
    & Bilinear$_{\text{sRGB}}$ & 31.73 & 0.846 & 0.225 & 28.94 & 0.894 & 0.135 & 27.65 & 0.897 & 0.168 & 25.90 & 0.861 & 0.096 \\
    & Bicubic$_{\text{sRGB}}$  & 31.63 & 0.842 & \underline{0.224} & \underline{29.05} & 0.898 & 0.121 & 27.65 & 0.897 & \underline{0.087} & 25.85 & 0.860 & 0.095 \\
    & Nearest$_{\text{sRGB}}$  & 30.99 & 0.828 & 0.232 & 28.34 & 0.882 & 0.132 & 26.12 & 0.857 & 0.095 & 24.23 & 0.802 & 0.109 \\
    & Area$_{\text{sRGB}}$     & \underline{31.89} & \underline{0.852} & 0.233 & 28.95 & 0.895 & 0.136 &\underline{27.90} & \underline{0.909} & 0.093 & 25.83 & 0.859 & 0.104 \\
    & CAR~\cite{CAR}           &    -  &   -   &   -   & 29.03 & \underline{0.906} & 0.155 &   -   &   -   &   -   & 26.00 & 0.882 & 0.120 \\
    &  HCD-IRN~\cite{HCD-IRN} &    -  &   -   &   -   & 29.00 & 0.896 & 0.133  &    -  &   -   &   - & \underline{27.05} & \underline{0.889} & 0.106 \\
    & IRN~\cite{IRN}           &    -  &   -   &   -   & 28.18 & \underline{0.906} & \underline{0.091} & 24.46 & 0.881 & 0.092 & 24.11 & 0.863 & \underline{0.091} \\
    & AIDN~\cite{AIDN}         &    -  &   -   &   -   & 20.37 & 0.687 & 0.330 & 19.06 & 0.583 & 0.458 & 19.14 & 0.602 & 0.377 \\
    & SDFlow~\cite{SDFlow} &    -  &   -   &   -   &   -  &   -   &   -   &   -  &   -   &   -   & 26.45 & 0.889 & 0.148 \\
    & TIRN~\cite{TSAIN-TIRN}   &    -  &   -   &   -   & 27.30 & 0.849 & 0.118 &   -   &   -   &   -   & 22.39 & 0.763 & 0.148 \\
    \midrule
    \multirow{5}[2]{*}{\rotatebox{90}{RAW-based}} 
    & Bilinear$_{\text{RAW}}$  & 31.85 & 0.851 & 0.231 & 27.18 & 0.876 & 0.130 & 24.50 & 0.849 & 0.131 & 23.89 & 0.827 & 0.147 \\
    & Bicubic$_{\text{RAW}}$   & 31.55 & 0.841 & 0.225 & 27.38 & 0.888 & 0.107 & 24.50 & 0.849 & 0.131 & 23.44 & 0.813 & 0.158 \\
    & Nearest$_{\text{RAW}}$   & 30.16 & 0.809 & 0.255 & 25.59 & 0.827 & 0.117 & 21.81 & 0.701 & 0.148 & 20.45 & 0.616 & 0.177 \\
    & Area$_{\text{RAW}}$      & 31.75 & 0.846 & 0.225 & 27.18 & 0.876 & 0.130 & 24.62 & 0.841 & 0.145 & 24.18 & 0.824 & 0.156 \\
    & Ours                     & \textbf{32.38} & \textbf{0.898} & \textbf{0.170} & \textbf{29.26} & \textbf{0.927} & \textbf{0.074} & \textbf{28.35} & \textbf{0.926} & \textbf{0.058} & \textbf{28.27} & \textbf{0.930} & \textbf{0.052} \\
    \bottomrule
    \end{tabular}}
    \caption{Quantitative comparisons in the sRGB domain of our method and comparison methods on our proposed Real-NIRD dataset with the downscaling factor of 1.3$\times$ and the Real-RawVSR~\cite{Real-rawvsr} dataset with the factor of 2$\times$, 3$\times$, and 4$\times$. For fair comparisons, we have retrained all learning-based methods using their officially released codes. We employ a fixed ISP pipeline for RAW-based methods to convert the reconstructed LR RAW images into sRGB format for evaluation. The best results are highlighted in \textbf{bold}, and the second-best results are in \underline{underlined}. `-' indicates the results are unavailable. }
  \label{tab: Quantitative_RGB}%
\end{table*}%

\textbf{Energy Maximization Loss.} Although energy sacrifice during downscaling is inevitable, the ideal situation is to preserve high-frequency energy, especially in texture and edge regions, while allowing energy loss primarily in homogeneous areas. Thus, we propose an energy maximization loss $\mathcal{L}_{em}$ to further enhance the preservation of sparse high-frequency information across three directions, promoting the predicted high-frequency information to retain energy levels similar to those in the high-resolution features, which can be formulated as:
\begin{equation}
    \mathcal{L}_{em} = \sum_{k=1}^{K} \sum_{C \in \{V, H, D\}} \frac{\operatorname{abs}(||\hat{C}^{k}_{LR}||_2-||C^{k}_{HR}||_2)}{h^{\prime}w^{\prime}},
\end{equation}
where $||\hat{C}^{k}_{LR}||_2$ denotes the energy of the predicted high-frequency components and $||C^{k}_{HR}||_2$ represents the corresponding high-frequency energy extracted from the HR RAW image through the $k$-level 2D-DWT process. By minimizing such discrepancy, the model would be guided to better preserve essential high-frequency information during the downscaling process.

\subsection{Network Training}
To ensure the structure consistency between predicted coefficients in the wavelet domain and avoid cumulative error during the progressive iteration, we design a hierarchical wavelet-components consistency loss $\mathcal{L}_{hwc}$ to constrain the predicted coefficients similar to ground-truth counterparts, which is formulated as,
\begin{equation}
    \mathcal{L}_{hwc} =\sum_{k=1}^{K}{\sum_{C\in {\{A,V,H,D\}}}||\hat{C}^{k}_{LR} - C^{k}_{LR}||_1},
\end{equation}
where $||\cdot||_1$ denotes the L1 loss. Moreover, we utilize a content consistency loss $\mathcal{L}_{con}$ to minimize the content difference between the predicted LR RAW image $\hat{I}_{LR}$ and the reference LR image ${I}_{LR}$ as: $\mathcal{L}_{con}=||\hat{I}_{LR} - {I}_{LR}||_1$. The overall objective function we adopted to optimize the network can be formulated as:
\begin{equation}
 \mathcal{L}_{total} = \mathcal{L}_{con} + \mathcal{L}_{hwc} + \lambda \mathcal{L}_{em}.
\end{equation}
\begin{figure*}[!t]
    \centering
    \includegraphics[width=\linewidth]{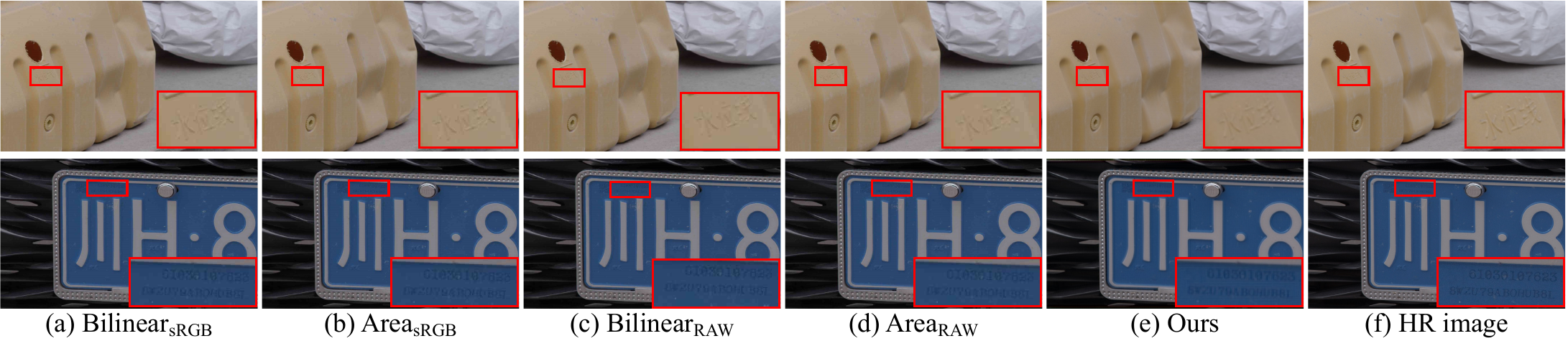}
    \caption{Qualitative comparison of our method with competitive methods Bilinear$_{\text{sRGB}}$, Area$_{\text{sRGB}}$, Bilinear$_{\text{RAW}}$ and Area$_{\text{RAW}}$ for 1.3$\times$ image downscaling. Error-prone regions are highlighted in red boxes. Best viewed by zooming in.}
    \label{fig: Qualitative Comparison_1.3x}
\end{figure*}

\begin{figure*}[!t]
        \centering
    \includegraphics[width=\linewidth]{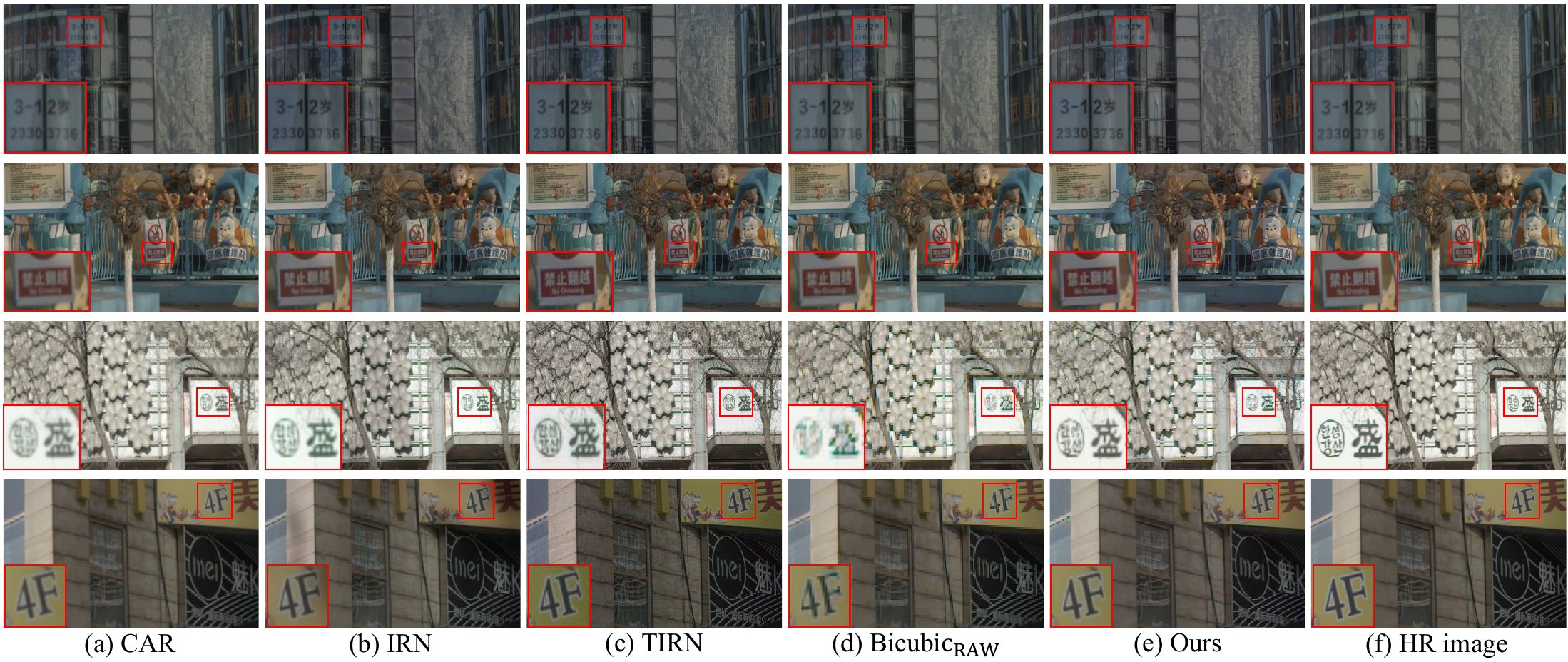}
    \caption{Qualitative comparison of our method with competitive methods CAR~\cite{CAR}, IRN~\cite{IRN}, TIRN~\cite{TSAIN-TIRN}, and Bicubic$_{\text{RAW}}$ for 2$\times$ (rows 1-2) and 4$\times$ (rows 3-4) scales. Error-prone regions are highlighted in red boxes. Best viewed by zooming in.}
    \label{fig: Qualitative Comparison_4x}
\end{figure*}

\section{Experiments}\label{sec: experiments}
\subsection{Dataset}\label{subsec: dataset}
\textbf{Real-RawVSR Dataset.} The Real-RawVSR dataset~\cite{Real-rawvsr} contains 450 video pairs captured with two Canon 60D cameras in various realistic indoor and outdoor scenes using a beam splitter to eliminate parallax, where the low-resolution (LR) and high-resolution (HR) pairs are aligned using the RANSAC algorithm~\cite{RANSAC} to correct for lens distortion field-of-view variance. The captured videos come in 2, 3, and 4 magnification factors, offering LR-HR pairs in both RAW and sRGB formats, and each of them contains approximately 150 frames. Considering there lacks a dedicated dataset specially designed for RAW image downscaling, we employ the HR images and LR images of the Real-RawVSR dataset as the input and the reference respectively for benchmarking previous methods and our approach. We retain 6,200 HR-LR RAW image pairs for training for each downsampling factor and randomly select 637, 542, and 549 pairs for evaluation at 2$\times$, 3$\times$, and 4$\times$, respectively.

\textbf{Real-NIRD Dataset.} Since the Real-RawVSR dataset only provides image pairs with integer scaling factors, we present the \textbf{Real}istic \textbf{N}on-\textbf{I}nteger \textbf{R}AW \textbf{D}ownsacling Dataset (Real-NIRD) that contains HR-LR image pairs with non-integer downscaling factor of 1.3$\times$ captured in realistic scenes using Nikon Z7 camera for benchmarking the non-integer image downscaling task. The original resolutions of HR images and LR RAW images are 8,256$\times$5,504 and 6,192$\times$4,128, respectively. To ensure content consistency and avoid the misalignment of the image pairs, we adopt tripods to mount the camera with remote applications for data collection. Moreover, we crop the original full-resolution images to improve the scene richness of the dataset, resulting in 2,000 image pairs for training and 149 pairs for evaluation. 

\subsection{Implementation Details}\label{subsec: implementation}
We implement the proposed method with PyTorch on eight NVIDIA RTX 2080Ti GPUs, where the batch size is set to 8. We employ the Adam optimizer~\cite{Adam} for optimization with the initial learning rate set to $1\times10^{-4}$ and decay by a factor of 0.8 for each $4\times10^4$ iterations, where the network can be converged for $1.6\times10^5$ iterations. The hyper-parameters $K$ and $\lambda$ are set to 4 and 0.1, respectively.

\begin{table}[!t]
  \centering
  \huge
    \resizebox{\linewidth}{!}{
    \begin{tabular}{c|l|ccccc}
    \toprule
     & Metric & Bilinear$_{\text{RAW}}$ & Bicubic$_{\text{RAW}}$ & Nearest$_{\text{RAW}}$ & Area$_{\text{RAW}}$  & Ours \\
    \midrule
    \multirow{2}[2]{*}{1.3$\times$} 
    & PSNR $\uparrow$ &  35.80 & 35.37 & 33.57 & \underline{35.95} & \textbf{40.39} \\
    & SSIM $\uparrow$ & 0.836 & 0.833 & 0.824 & \underline{0.837} & \textbf{0.984} \\
    \midrule
    \multirow{2}[2]{*}{2$\times$}
    & PSNR $\uparrow$ & 35.36 & \underline{35.52} & 33.55 & 35.36 & \textbf{36.76} \\
    & SSIM $\uparrow$ & 0.952 & \underline{0.954} & 0.937 & 0.952 & \textbf{0.958} \\
    \midrule
    \multirow{2}[2]{*}{3$\times$} 
    & PSNR $\uparrow$ & 32.81 & 32.81 & 29.93 & \underline{32.88} & \textbf{35.66} \\
    & SSIM $\uparrow$ & \underline{0.932} & \underline{0.932} & 0.886 & 0.931 & \textbf{0.948} \\
    \midrule
    \multirow{2}[2]{*}{4$\times$} 
    & PSNR $\uparrow$ & 32.66 & 32.22 & 28.81 & \underline{32.99} & \textbf{35.75} \\
    & SSIM $\uparrow$ & 0.929 & 0.924 & 0.852 & \underline{0.931} & \textbf{0.954} \\
    \bottomrule
    \end{tabular}}
    \caption{Quantitative comparisons in the RAW domain of our method and traditional interpolation methods on the Real-NIRD and Real-RawVSR~\cite{Real-rawvsr} datasets. The best results are highlighted in \textbf{bold}, and the second-best are in \underline{underlined}.}
  \label{tab: Quantitative_RAW}
\end{table}%

\subsection{Comparison with Existing Method}
\textbf{Comparison Method.} In this section, we compare the proposed method with traditional interpolation methods including Nearest, Bilinear, Area, and Bicubic which can be applied to both sRGB and RAW image downscaling denoted with $_{\text{sRGB}}$ and $_{\text{RAW}}$, as well as learning-based sRGB image downscaling methods including CAR~\cite{CAR}, HCD-IRN~\cite{HCD-IRN}, IRN~\cite{IRN}, AIDN~\cite{AIDN}, SDFlow~\cite{SDFlow} and TIRN~\cite{TSAIN-TIRN}. For fair comparisons, we use a fixed ISP pipeline to convert the HR and LR RAW images in our Real-NIRD dataset and the Real-RawVSR~\cite{Real-rawvsr} dataset into sRGB images to retrain the learning-based sRGB downscaling methods. For quantitative evaluation, we conduct experiments on the sRGB and RAW space separately, where the results of our method and the RAW-based interpolation methods are also converted to sRGB images using the ISP pipeline for evaluation in sRGB space. Two distortion metrics PSNR and SSIM~\cite{SSIM}, and a perceptual metric LPIPS~\cite{LPIPS} are adopted for evaluation.

\textbf{Quantitative Comparison.} We evaluate our method and comparison methods in the sRGB domain using the Real-NIRD dataset with a 1.3$\times$ non-integer downscaling factor and the Real-RawVSR~\cite{Real-rawvsr} dataset with integer factors (2$\times$, 3$\times$, 4$\times$). As shown in Table~\ref{tab: Quantitative_RGB}, previous learning-based sRGB downscaling methods even underperform traditional interpolation techniques, since their primary target is not focusing on LR image reconstruction. Moreover, traditional interpolation methods conducted in the RAW space present noticeable inferiority to the sRGB domain under integer settings, due to RAW’s nonlinearity and high dynamic range, making it more prone to distortion. In contrast, our method achieves state-of-the-art performance across all settings, benefiting from wavelet transformation and the well-designed new components. Notably, our method significantly outperforms existing methods, particularly under the non-integer scaling factor of 1.3$\times$ and the larger integer factor of 4$\times$. Specifically, for the 4$\times$ setting, our method achieves improvements of 0.53 in PSNR, 0.016 in SSIM, and 0.021 in LPIPS, while the performance gains are even more pronounced under the 1.3$\times$ configuration, with improvements of 0.61 in PSNR, 0.055 in SSIM, and 0.066 in LPIPS, highlighting the ability to generate visually satisfactory results and suitability for the RAW image downscaling task. Furthermore, we also evaluate the performance of our method and traditional interpolation methods in the RAW domain as reported in Table~\ref{tab: Quantitative_RAW}, where our method presents noticeable performance gains across all settings, further confirming its ability to generate high-fidelity low-resolution RAW images.

\textbf{Qualitative Comparison.} We present visual comparisons of our method and competitive methods on the proposed Real-NIRD dataset in Fig.~\ref{fig: Qualitative Comparison_1.3x} and the Real-RawVSR~\cite{Real-rawvsr} dataset for 2$\times$ and 4$\times$ downscaling in Fig.~\ref{fig: Qualitative Comparison_4x}, where the reconstructed LR RAW images of our method and RAW-based interpolation methods are visualized by a fixed ISP pipeline. Previous learning-based sRGB downscaling methods fail to maintain detail sharpness, whereas IRN even produces black ghosting artifacts that make content inconsistent with reference HR images. Traditional interpolation methods applied directly to RAW data produce pronounced jagged edges, severely compromising edge information, particularly at large downsampling ratios, i.e., 4$\times$. In contrast, our method achieves more detailed, visually appealing reconstructions, and exceptional visual fidelity to the high-quality HR images, proving its effectiveness.
\subsection{Ablation Study}
In this section, we conduct a series of ablation studies to validate the effectiveness of the newly proposed component in our method. We use the implementation details described in Sec.~\ref{subsec: implementation} for training and quantitative results in the sRGB domain on the Real-RawVSR~\cite{Real-rawvsr} dataset with the downscaling factor of  2$\times$ are reported in Table~\ref{tab: Ablation}. 

\begin{figure}[!t]
    \centering
    \includegraphics[width=\linewidth]{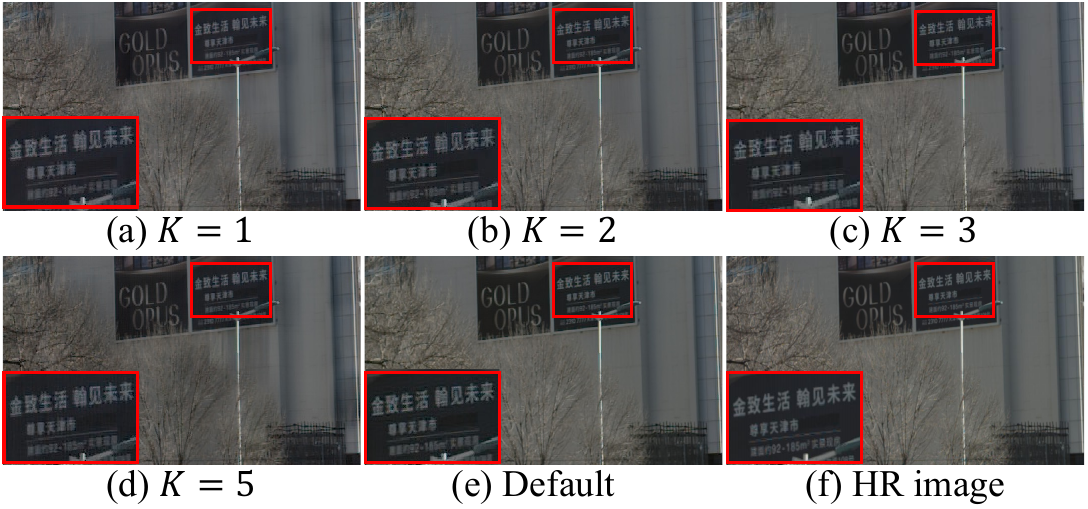}
    \caption{Visual results of the ablation study about our recurrent reconstruction framework. Best viewed by zooming in.}
    \label{fig: ablation_study1}
\end{figure}

\textbf{Recurrent Reconstruction Framework.} To validate the impact of recurrent times $K$ on performance, we systematically vary $K$ from 1 to 5. As reported in rows 1-4 of Table~\ref{tab: Ablation}, the overall performance improves as $K$ gradually increases, but starts to degrade at $K=5$
due to reduced information richness in the encoded feature $f_K$ at excessive iterations, hindering accurate high-frequency prediction. Visual comparisons in Fig.~\ref{fig: ablation_study1} demonstrate that increasing $K$ from 1 to 4 progressively improves high-frequency details, especially edge sharpness and texture. However, the $K=5$ configuration shown in Fig.~\ref{fig: ablation_study1} (d) exhibits visible high-frequency detail loss compared to the default setting, i.e., $K=4$.

\begin{table}[!t]
    \centering
    \resizebox{0.6\linewidth}{!}{
    \begin{tabular}{l|ccc}
    \toprule
         Method & PSNR $\uparrow$ & SSIM $\uparrow$ & LPIPS $\downarrow$\\
         \midrule
        $K=1$ & 27.12 & 0.889 & 0.112 \\
        $K=2$ & 27.30 & 0.898 & 0.100 \\
        $K=3$ & 28.08 & 0.904 & 0.099 \\
        $K=5$ & 26.91 & 0.881 & 0.113 \\
        \midrule
        w/o LASDM & 28.30 & 0.903 & 0.101 \\
        w/o HFPM & 28.44 & 0.903 & 0.102 \\
        w/o $\mathcal{L}_{em}$ & 28.99 & 0.909 & 0.090 \\
        \midrule
        Default & 29.26 & 0.927 & 0.074 \\
     \bottomrule
    \end{tabular}}
    \caption{Ablation studies of the recurrent times $K$, the newly proposed modules, and loss function, please refer to the text for more details. `w/o’ denotes without.}
    \label{tab: Ablation}
\end{table}

\begin{figure}[!t]
    \centering
    \includegraphics[width=\linewidth]{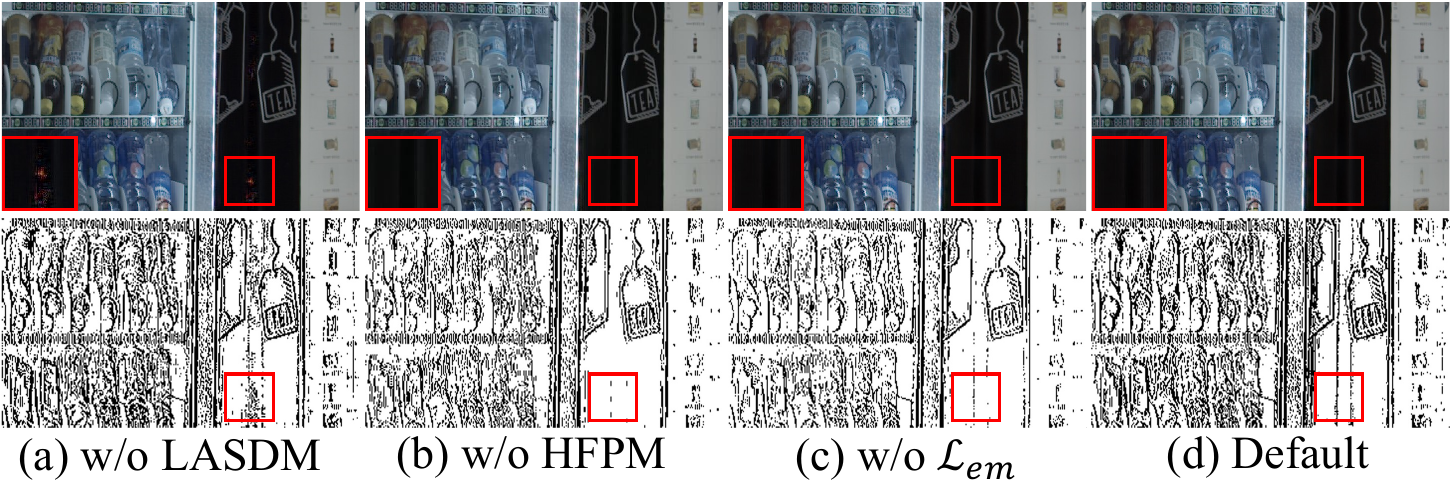}
    \caption{Visual results of the ablation study about our newly proposed modules and loss function. The high-frequency vertical wavelet components are shown below.}
    \label{fig: ablation_study2}
\end{figure}

\textbf{Low-Frequency Arbitrary-Scale Downscaling Module.} To evaluate the effectiveness of the proposed Low-Frequency Arbitrary-Scale Downscaling Module (LASDM), we replace it with traditional interpolation methods Bilinear while maintaining an identical CNN architecture to generate the initial average coefficient. As reported in row 5 and row 8 of Table~\ref{tab: Ablation}, our LASDM obtains performance gains in all metrics, benefiting from effectively utilizing high-resolution features to yield finer detail representation, which demonstrates that our LASDM is more suitable for the RAW image downscaling task. As shown in Fig.~\ref{fig: ablation_study2} (a) and (d), traditional interpolation approaches lead to unexpected artifacts, whereas our LASDM works well in preserving low-frequency information.

\textbf{High-Frequency Prediction Module.} To validate the effectiveness of our proposed High-Frequency Prediction Module (HFPM), we conduct an experiment by substituting it with a CNN module followed by bilinear interpolation for low-resolution high-frequency coefficient generation. As reported in row 6 and row 8 of Table~\ref{tab: Ablation}, incorporating the HFPM yields overall improvements across evaluation metrics, demonstrating that our HFPM is capable of accurately predicting the initial high-frequency coefficient. Fig.~\ref{fig: ablation_study2} (b) and (d) present discernible texture attenuation in high-frequency regions in the absence of our proposed HFPM, confirming its critical role in maintaining high-frequency fidelity.

\textbf{Loss Function.} To validate the effectiveness of our energy maximization loss function $\mathcal{L}_{em}$, we conduct an ablation study by removing it and retraining the network. As shown in rows 7-8 of Table~\ref{tab: Ablation}, incorporating $\mathcal{L}_{em}$ noticeably boosts performance across all metrics, confirming its role in accurately predicting high-frequency coefficients. Furthermore, Fig.~\ref{fig: ablation_study2} (c) and (d) illustrate that omitting the $\mathcal{L}_{em}$ leads to texture loss in high-frequency components, highlighting its contribution to generating finer details.

\section{Conclusion}
\label{sec: conclusion} 
We have presented a recurrent reconstruction framework for the arbitrary-scale RAW image downscaling. Technically, we introduce a wavelet-based recurrent reconstruction decoder that iteratively reconstructs the low-resolution RAW image, where an energy maximization loss is further proposed to refine the predicted high-frequency coefficients. Meanwhile, we present a low-frequency arbitrary-scale downscaling module to derive the average coefficient for arbitrary-scale downscaling, and a high-frequency prediction module to predict the low-resolution high-frequency components from high-resolution high-frequency features and the predicted target low-frequency coefficient. Furthermore, we introduce the Real-NIRD dataset with a 1.3$\times$ non-integer downscaling factor, alongside publicly available datasets for benchmarking. Extensive experiments show that our method outperforms state-of-the-art competitors both quantitatively and visually.


\begin{acks}
This work was supported in part by the National Natural Science Foundation of China (NSFC) under Grant Nos. 62271334, 62372091, Hainan Province Key R\&D Program under Grant No. ZDYF2024(LALH)001 and the 173 Program under Grant No.2020-JCJQ-ZD-060-00.
\end{acks}
\bibliographystyle{ACM-Reference-Format}
\bibliography{sample-base}


\end{document}